\documentclass[twocolumn,twoside,slac_two]{revtex4}
\usepackage{graphicx}
\usepackage{fancyhdr}
\newcommand{\hess}{\textsc{H.E.S.S.}}

\begin{document}

\title{The Extragalactic sky with the High Energy Stereoscopic System.}

%

\author{D.~A.~Sanchez}
\affiliation{Laboratoire d'Annecy-le-Vieux de Physique des Particules, Universit\'e de Savoie, CNRS/IN2P3}
\author{On behalf of the \hess\ collaboration}

\begin{abstract}

The number of extragalactic sources detected at very hight energy (VHE,
E$>$100GeV) has dramatically increased during the past years to reach more than
fifty. The High Energy Stereoscopic System (H.E.S.S.) had observed the sky for
more than 10 years now and discovered about twenty objects. With the advent of
the fifth 28 meters telescope, the H.E.S.S. energy range extends down to ~30
GeV. When H.E.S.S. data are combined with the data of the Fermi Large area
Telescope, the covered energy range is of several decades allowing an
unprecedented description of the spectrum of extragalactic objects. In this
talk, a review of the extragalactic sources studied with H.E.S.S. will be given
together with first H.E.S.S. phase II results on extragalactic sources.

\end{abstract}

\maketitle

\thispagestyle{fancy}

\section{The H.E.S.S. array}

The High Energy Stereoscopic System (H.E.S.S.) is located near the Gamsberg
mountain in Namibia at an altitude of 1800 meters. H.E.S.S. detects $\gamma$-ray
photons by recording the Cherenkov light produced by the electromagnetic shower
resulting from the interaction of the photons with the atmosphere.  

The Phase I of the project was completed in December 2003. At this time the
array was made of four 12-meters telescopes. Each telescope has a camera
composed of 960 Photo-multipliers (PMTs) and works in stereoscopic mode. 

\begin{figure}
\includegraphics[width=85mm]{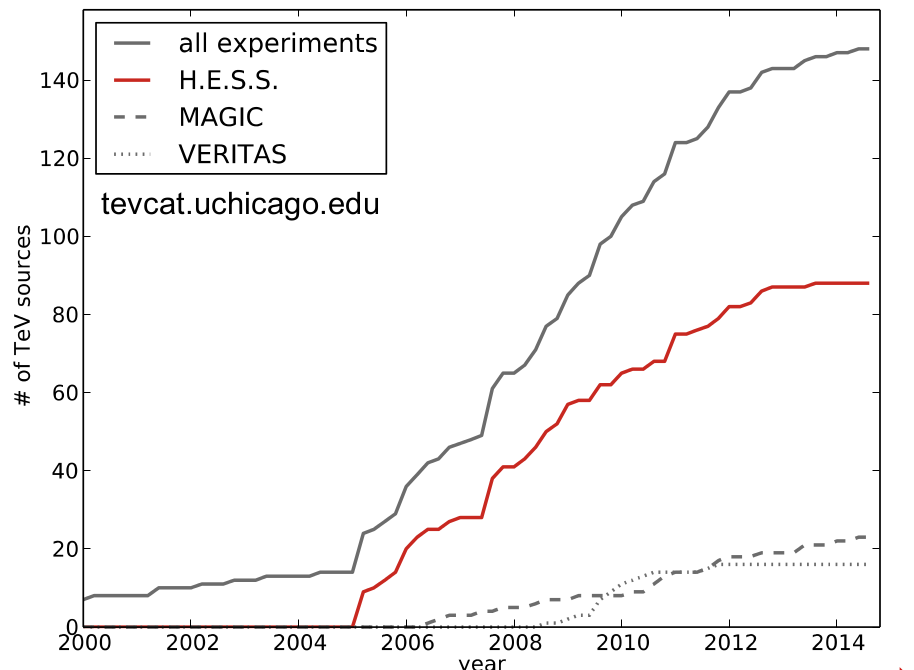}
\caption{Extragalactic and Galactic sources detected in the VHE range. The gray line is the total of discovered object while the red one is the contribution of the H.E.S.S. experiment.}
\label{source}
\end{figure}

The H.E.S.S. Phase I array has a field of view of 5 degrees, an angular
resolution of 0.1 degree for an energy threshold down to 100~GeV. The array is
taking data for more than 10 years now and has increased the catalogue of
sources detected in the very high energy range (VHE, $E>$100~GeV) and our
knowledge of the field. H.E.S.S. has discovered more than 80 objects (galactic
or extragalactic, Fig. \ref{source}), performed a deep Galactic plane survey,
dark matter searches, multiwavelength campaigns with other instruments and
studies of the extragalactic background light with blazars. 

One of the aims of H.E.S.S. is the detection of gamma-ray bursts (GRBs). While no
GRB has been detected so far in the VHE range, more than 20 follow-up have been
performed \citep{grbs}. H.E.S.S. also monitors variable and bright objects and responds
to target of opportunity in order to better know the mechanisms that produce the
variability of blazars.

\begin{figure}
\includegraphics[width=85mm]{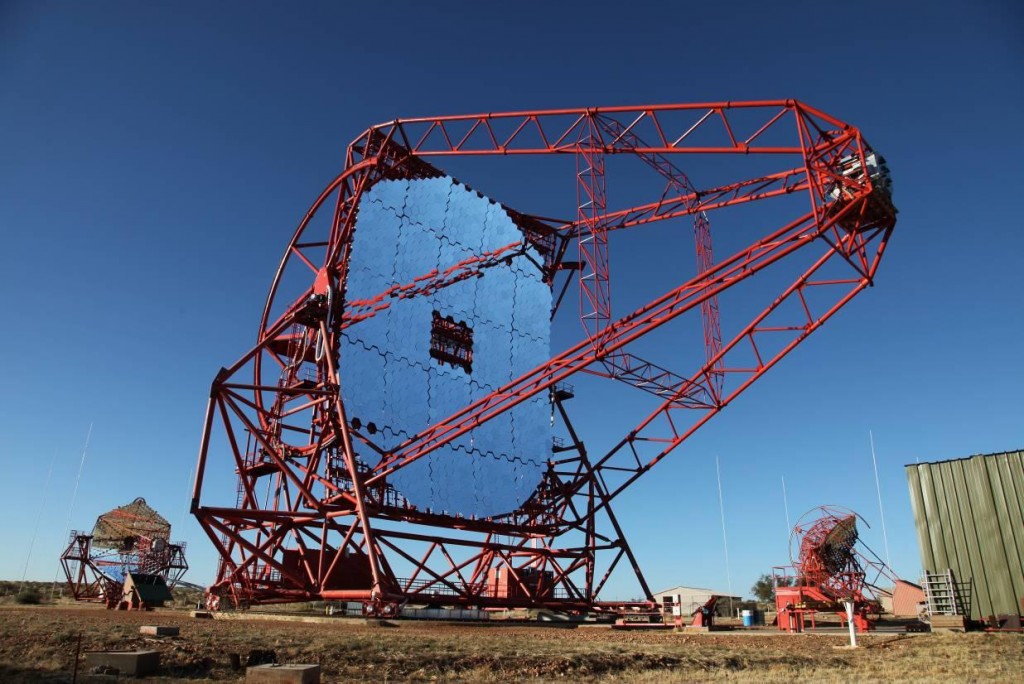}
\caption{The fifth H.E.S.S. telescope.}
\label{CT5}
\end{figure}


The experiment entered into its Phase II with the addition of a fifth telescope
(named CT5) placed in the middle of the array. The dish is 32.6 meters by 24.3
meters, equivalent to 28-meters circular dish for a focal length of 36 meters.
The telescope is equipped with a Alt-Az mount \cite{drive}. The camera composed
of 2048 PMTs for a total weight of $\approx$ 3 tons, can record 3600 images per
second \cite{camera} and is mounted on an auto-focus system. A picture of this
telescope is show on figure \ref{CT5}. The field of view of this telescope is
3.5 degrees for an angular resolution from $\approx$ 0.4 degree to less than
0.1. The energy coverage of the H.E.S.S. experiment is then extended down to
energies of a few tens of GeV. Characteristics of the CT5 telescope are
summarized in Table \ref{tablect5}.

H.E.S.S. Phase II is the first hybrid array of Cherenkov telescopes and is
designed to work in different configurations. Data can be taken by CT5 only in
the so-called \textit{Mono mode}. The \textit{hybrid mode} involves all five
telescopes for a better sensitivity in the entire energy range. Stereoscopic
observations with only the four 12-meters telescopes are still possible. The
ability to split the array in 2 (CT5 Mono mode + the four 12 meters telescopes)
allows to increase the observation time which is rather low for such an
experiment ($\approx$ 1000 hours per year). 

\begin{table}[t]
\begin{center}
\caption{Characteristics of the CT5 telescope.}
  \begin{tabular}{|l|c|}\hline
 Mount type &	Alt-Az mount \\
Height of elevation axis &	24 m	\\\hline
Dish Dimensions &	32.6 m by 24.3 m\\
& equivalent to 28 m circular dish\\
Focal length&	36 m\\
Total mirror area	&614 m$^2$\\\hline
Photo sensors	&2048\\
Pixel size&	42 mm\\ 
FoV & 3.5 degrees \\
Camera weight	&3.0 tons\\
\hline
  \end{tabular}%
\label{tablect5}
\end{center}
\end{table}

In this work, the last results of the H.E.S.S. Phase I array are presented and
new preliminary results of observations carry in CT5 Mono mode on extragalactic
targets are given.

\section{Recent H.E.S.S. phase I results}
\subsection{Long term monitoring of PKS~2155-304}

The high frequency peaked BL Lac (HBL) object PKS 2155-304 is the brightest
object of the southern sky in VHE. This source has been the target of several campaigns
in the past involving H.E.S.S. \cite{2155mwl1,2155mwl2} and other instruments.
This HBL is also famous for the flare that happened in June 2006 with variation at
the minute time scale \cite{flare}. 

Due to its brightness, the source has been monitored by the H.E.S.S. telescopes
since 2004. Data taken between 2004 and 2012, except the June/July exposures where the
sources underwent a flare, were analysed using a Hillas-type analysis
\cite{crab}. The nightly binned light curve above 300~GeV has been used for the
analysis and the corresponding mean flux is $2.02 \times
10^{-11}$cm$^{-2}$\,s$^{-1}$.

\begin{figure}[Ht!]
     \includegraphics[width=80mm]{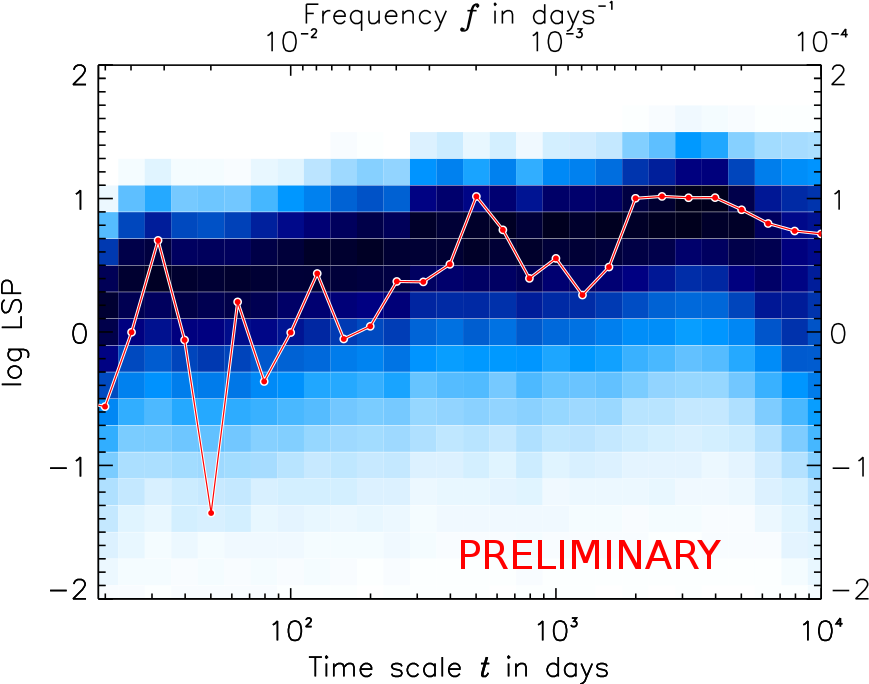}
\caption{Power Spectral density of PKS 2155-304 (red line and dots) obtained with the H.E.S.S. data. The color area gives the results of the simulations used to fit the data.}
\label{PSD}
 \end{figure}

The power spectral density (PSD) has been calculated (Fig \ref{PSD}) and fitted with a power law of index $\beta$. A forward folding method together with a likelihood estimator described in \cite{max} were used to fit the model to the data. The best fit value is $\beta = 0.9\pm 0.2$ for the data set presented here which corresponds to the source being at a low flux state. This has to be compared with the value found during the 2006 flare $\beta=2$ but on short time-scales. This may be a sign of a break in the PSD  with a change of $\beta$ from 2 to 1 or a change in the PSD between the two flux states. 

\subsection{The 2012 Flare of PG 1553+113}

In March 2012, the HBL  PG 1553+113 underwent a flare observed by \hess\ during 2 nights. This data set have been used to determine the source redshift and possible Lorentz invariance violation (LIV) effects. The redshift has been constrained to be $z = 0.49 \pm 0.04$ and an lower limit on the energy scale at which LIV effects take place has been set to $\textrm{E}_{\rm QG,1}>4.10\times 10^{17}$~GeV and 
$\textrm{E}_{\rm QG,2}>2.10\times 10^{10}$~GeV for linear and quadratic LIV effects. More details can be found in \cite{proc1553} and \cite{paper1553}

\section{H.E.S.S. II first results on extragalactic objects}

Since the inauguration on September 2012, CT5 has been in a commissioning phase. The HBLs PKS~2155-304 and PG 1553+113 serve as calibration targets. The analysis of the data was performed using the Model analysis \cite{model} with cuts adapted for the Mono observations with CT5.

\paragraph{PG 1553+113}

\begin{table}[t]
\begin{center}
\caption{Preliminary results on PG 1553+113 obtained with CT5 in 2014.}
  \begin{tabular}{|l|c|}\hline
    Live Time &  15.1 h\\\hline
    Excess & 2508 $\gamma$ \\\hline
    Significance & 26.6 $\sigma$ \\\hline
    Zenith & $\approx 35^\circ$ \\ \hline
    Rate & $2.77\pm0.11\gamma$/min \\ \hline
  \end{tabular}%
\label{table1553}
\end{center}
\end{table}

\begin{figure*}[Hbt!]
  \begin{minipage}{0.9\textwidth}
     \includegraphics[width=65mm]{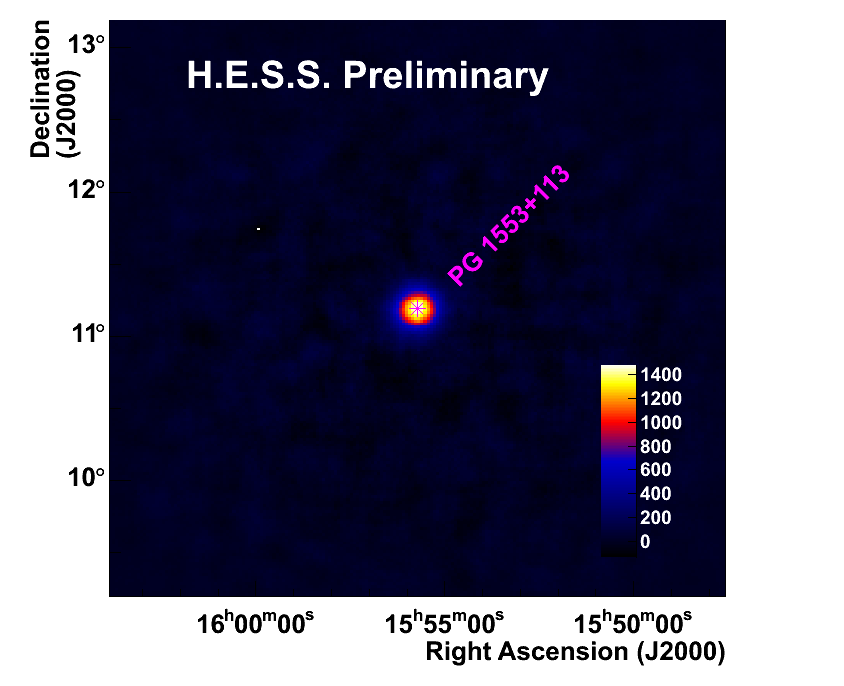}
     \includegraphics[width=85mm]{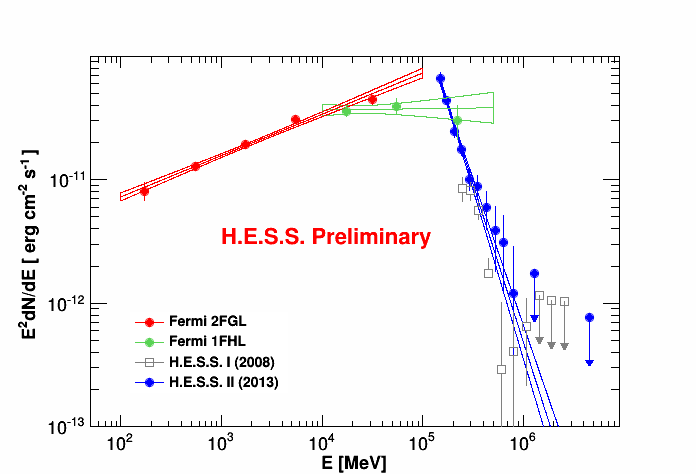}
\caption{Excess map (left) and spectral energy distribution (right) of PG 1553+113 obtained with CT5 in Mono mode. The CT5 SED is given by the blue points and contour. \textit{Fermi} 2FLG (red) and 1FHL (blue) results are also presented. The gray points are archival data from \cite{1553vlt}.}
\label{pg}
\end{minipage}

 \end{figure*}
This source has a soft spectrum in the VHE energy range, H.E.S.S. measured a spectral index of $\gamma=4.6 \pm 0.6$ \cite{1553vlt}. This makes it well suited for observations with CT5. A total of 15.1 hours of live time has been analysed and an excess of more than 2500 events has been found. This corresponds to $2.77\pm 0.11 \gamma$ per minutes for a significance of 26.6 sigma (Table \ref{table1553}). 

An excess map around the coordinates of the object and the resulting spectral energy distribution of the source are given on figure \ref{pg}. The spectrum measured with CT5 is compared with the non contemporaneous data from the \textit{Fermi} second source catalogue  (2FGL) \cite{fgl} and first high energy catalogue (1FHL) \cite{fhl}.

\paragraph{PKS 2155-304}

\begin{table}[t]
\begin{center}
\caption{Preliminary results on PKS 2155-304 obtained with CT5 in 2014.}
  \begin{tabular}{|l|c|}\hline
    Live Time & 42.9 h\\\hline
    Excess & 4442 $\gamma$\\\hline
    Significance & 29.7 $\sigma$ \\\hline
    Zenith & $\approx 21^\circ$ \\ \hline
    Rate & $1.72\pm0.06\gamma$/min \\ \hline
  \end{tabular}%
\label{table2155}
\end{center}
\end{table}

\begin{figure*}[Hbt!]
  \begin{minipage}{0.9\textwidth}
     \includegraphics[width=65mm]{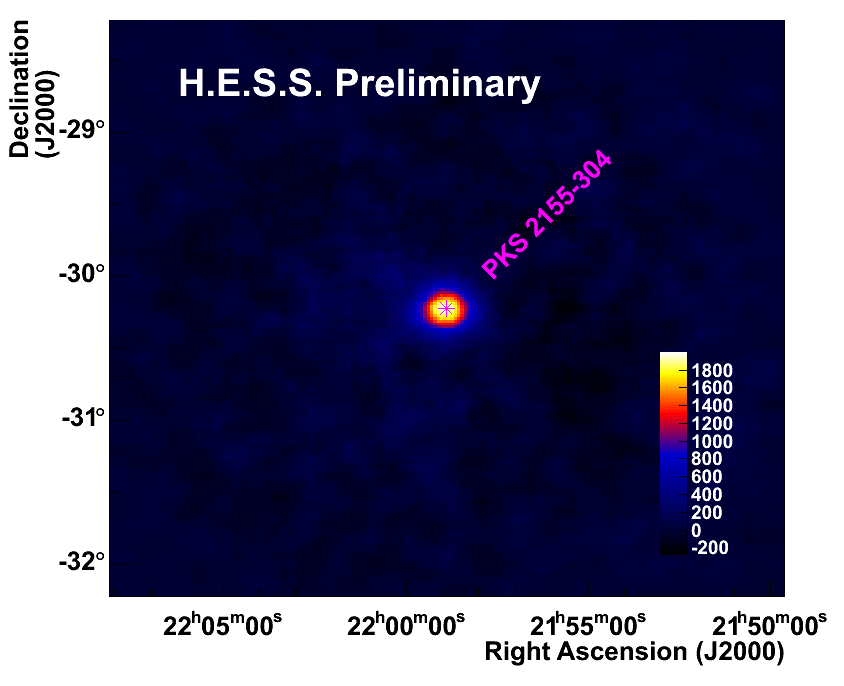}
     \includegraphics[width=85mm]{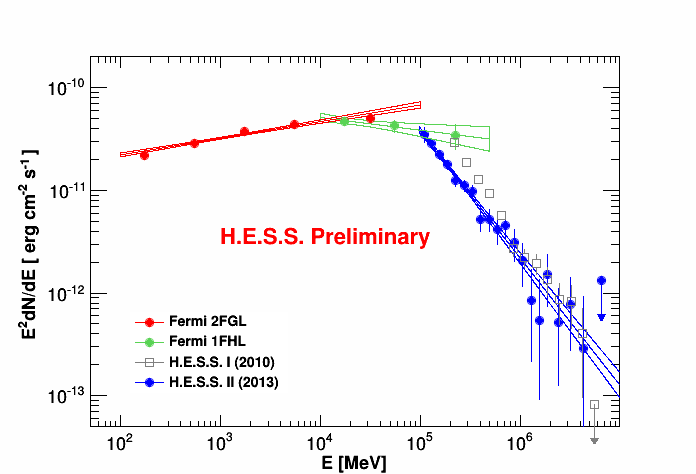}
\caption{Excess Map (left) and spectral energy distribution (right) of PKS 2155-304 obtained with CT5 in Mono mode. The CT5 SED is given by the blue points and contour. \textit{Fermi} 2FLG (red) and 1FHL (blue) results are also presented. The gray points are archival data from \cite{2155data}.}
\label{pksmap}
\end{minipage}

 \end{figure*}

PKS 2155-304 is naturally a good target for calibration purposes and  has been observed for a total live time of 42.9 hours in 2013. The analysis yields an excess of more the 4400 events for a detection at a 29.7 $\sigma$ level (Table \ref{table2155}). Excess map of the field of view and SED are shown on figure \ref{pksmap}. The 2FGL and 1FHL are also used for comparison. 

A large part of the data were taken during a multiwavelength campaign with NuSTAR and \textit{Fermi}. These data were used to build the most precise SED of this object to date:  NuSTAR extended the X-ray spectrum up to 79 keV and \textit{Fermi} PASS 8 data were used. More details and results are given in \cite{proc2155} and in a forthcoming publication.

\section{CT5 as a transients machine}

\begin{figure}[Ht!]
     \includegraphics[width=80mm]{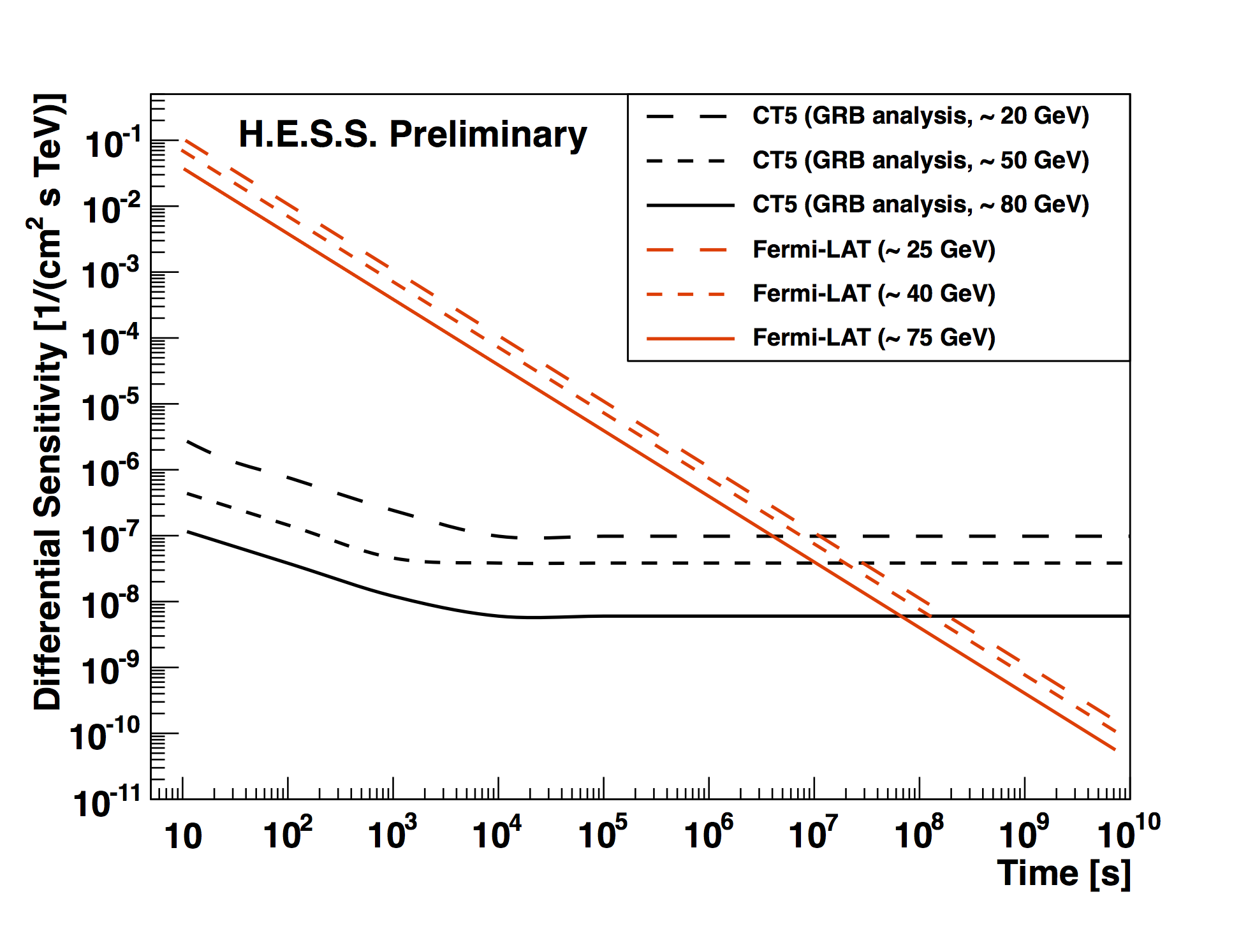}
\caption{Comparison of the differential sensitivity of the \textit{Fermi}-LAT and CT5 as a function of time and for different energy thresholds. }
\label{Sens}
 \end{figure}

One of the main goals of the H.E.S.S. Phase II is to study the variability on short time-scales. Figure \ref{Sens} presents the  differential sensitivity of CT5 and \textit{Fermi} as a function of time.  Below $10^7$ seconds, CT5 is clearly much more sensitive that the LAT above 25 GeV. This opens a new window for the detection of GRBs in this energy range. An alert system has been developed to reply to GCN alerts and an automatic re-pointing procedure is in place in case of such an alert. CT5 can, in such case, be on target within a minute \cite{drive} allowing prompt observations.

\bigskip 

\section{Conclusions}

The H.E.S.S. array is taking data now since more than 10 years and has allowed many discoveries. With the advent of the Phase II of the experiment, a new window has been opened in the 30 GeV-100 GeV energy range. After a commissioning phase, CT5 is now running running in a normal operation mode.  

\section*{Acknowledgments} \label{Ack}

The support of the Namibian authorities and of the University of Namibia in facilitating the construction and operation of H.E.S.S. is gratefully acknowledged, as is the support by the German Ministry for Education and Research
(BMBF), the Max Planck Society, the French Ministry for Research, the CNRS-IN2P3 and the Astroparticle Interdisciplinary Programme of the CNRS, the U.K. Science and Technology Facilities Council (STFC), the
IPNP of the Charles University, the Polish Ministry of Science and Higher Education, the South African Department of Science and Technology and National Research Foundation, and by the University of Namibia. We appreciate
the excellent work of the technical support staff in Berlin, Durham, Hamburg, Heidelberg, Palaiseau, Paris, Saclay, and in Namibia in the construction and operation of the equipment.

The \textit{Fermi}-LAT Collaboration acknowledges support for LAT development, operation and data analysis from NASA and DOE (United States), CEA/Irfu and IN2P3/CNRS (France), ASI and INFN (Italy), MEXT, KEK, and JAXA (Japan), and the K.A.~Wallenberg Foundation, the Swedish Research Council and the National Space Board (Sweden). Science analysis support in the operations phase from INAF (Italy) and CNES (France) is also gratefully acknowledged.

The work of DS has been supported by the Investissements d'avenir, Labex ENIGMASS.

\bigskip 

\begin{thebibliography}{9}   






\bibitem{2155mwl1} Aharonian, F., Akhperjanian, A.~G., Bazer-Bachi, A.~R., et al.\ 2005, A\&A, 442, 895 
\bibitem{1553vlt} Aharonian, F., Akhperjanian, A.~G., Barres de Almeida, U., et al.\ 2008, A\&A, 477, 481 
\bibitem{grbs} Aharonian, F., Akhperjanian, A.~G., Barres de Almeida, U., et al.\ 2009, A\&A, 495, 505 
\bibitem{2155mwl2} Aharonian, F., Akhperjanian, A.~G., Anton, G., et al.\ 2009, ApJL, 696, L150 
\bibitem{2155data} Abramowski, A., Acero, F., et al.\ 2010, A\&A, 520, AA83 
\bibitem{flare} Aharonian, F., et al.\ 2007, ApJL, 664, L71 
\bibitem{crab} Aharonian, F., et al.\ 2006, A\&A, 457, 899 
\bibitem{paper1553}  Abramowski, A., Aharonian, F., Ait Benkhali, F., et al.\ 2015, arXiv:1501.05087 
\bibitem{camera} Bolmont, J., Corona, P., Gauron, P., et al.\ 2014, Nuclear Instruments and Methods in Physics Research A, 761, 46
\bibitem{model} de Naurois, M. \& Rolland, L. 2009, Astroparticle Physics, 32, 231
\bibitem{drive} Hofverberg, P., Kankanyan, R., Panter, M., et al.\ 2013, arXiv:1307.4550 
\bibitem{max} Kastendieck, M.~A., Ashley, M.~C.~B., \& Horns, D.\ 2011, A\&A, 531, AA123 
\bibitem{fgl} Nolan, P.~L., Abdo, A.~A., Ackermann, M., et al.\ 2012, ApJS, 199, 31 
\bibitem{proc1553} Sanchez D.A. et al, Probe of Lorentz Invariance Violation effects and determination of the distance of PG 1553+113, these proceedings. 
\bibitem{proc2155} Sanchez D.A. et al, Multiwavelength campaign on the HBL PKS~2155-304: A new insight on its spectral energy distribution, these proceedings. 

\bibitem{fhl} The Fermi-LAT Collaboration 2013, arXiv:1306.6772 



\end{thebibliography}

\end{document}